\documentclass[pre,twocolumn,superscriptaddress,floatfix]{revtex4-1}  %floatfix

\usepackage[paperwidth=210mm,paperheight=297mm,centering,hmargin=2cm,vmargin=2.5cm]{geometry}

\usepackage[pdftex]{graphicx}
\usepackage{amsmath,amsfonts,amssymb}
\usepackage{natbib}
\usepackage{MnSymbol}
\usepackage{csquotes}

\usepackage{hyperref}
\usepackage{xcolor} 
\usepackage{array}
\usepackage{pgfplots}
\usepackage{graphicx}
\usepackage[left,modulo,pagewise]{lineno}
\usepackage{dsfont}

\definecolor{bluemoi}{rgb}{0.25,0.50 ,0.75} 

\hypersetup{
  bookmarksopen=true,
  pdftitle=" ",
  pdfauthor=" ", 
  pdfsubject=" ", 
  pdftoolbar=true, % toolbar hidden
  pdfmenubar=true, %menubar shown
  pdfhighlight=/O, %effect of clicking on a link
  colorlinks=true, %couleurs sur les liens hypertextes
  pdfpagemode=UseNone, %aucun mode de page
  pdfpagelayout=SinglePage, %ouverture en simple page
  pdffitwindow=true, %pages ouvertes entierement dans toute la fenetre
  linkcolor=bluemoi, %couleur des liens hypertextes internes
  citecolor=bluemoi, %couleur des liens pour les citations
  urlcolor=bluemoi %couleur des liens pour les url
}

\makeatletter
\renewcommand{\figurename}{\sf \textbf{Figure}}
\renewcommand{\thefigure}{\arabic{figure}}
\renewcommand{\fnum@figure}{\sf\textbf{\figurename}~\textbf{\thefigure}}
\renewcommand{\tablename}{\sf\textbf{Table}}
\renewcommand{\thetable}{\arabic{table}}
\renewcommand{\fnum@table}{\sf\textbf{\tablename}~\textbf{\thetable}}
\makeatother

\begin{document}

\title{Animal daily mobility patterns analysis using resting event networks} 

%\title{A resting event network approach to unveil animal daily mobility patterns} 

\author{Maxime Lenormand}
\thanks{Corresponding author: maxime.lenormand@inrae.fr}
\affiliation{TETIS, Univ Montpellier, AgroParisTech, Cirad, CNRS, INRAE, Montpellier, France}

\author{Herv\'e Pella}
\affiliation{INRAE, RiverLy, 5 rue de la Doua, CS 20244, 69625 Villeurbanne Cedex, France}

\author{Herv\'e Capra}
\affiliation{INRAE, RiverLy, 5 rue de la Doua, CS 20244, 69625 Villeurbanne Cedex, France}

\begin{abstract} 
Characterizing the movement patterns of animals is crucial to improve our understanding of their behavior and thus develop adequate conservation strategies. Such investigations, which could not have been implemented in practice only a few years ago, have been facilitated through the recent advances in tracking methods that enable researchers to study animal movement at an unprecedented spatio-temporal resolution. However, the identification and extraction of patterns from spatio-temporal trajectories is still a general problem that has relevance for many applications. Here, we rely on the concept of resting event networks to identify the presence of daily mobility patterns in animal spatio-temporal trajectories. We illustrate our approach by analyzing spatio-temporal trajectories of several fish species in a large hydropeaking river.
\end{abstract}

\maketitle

\begin{figure*}[!ht]
	\centering 
	\includegraphics[width=14cm]{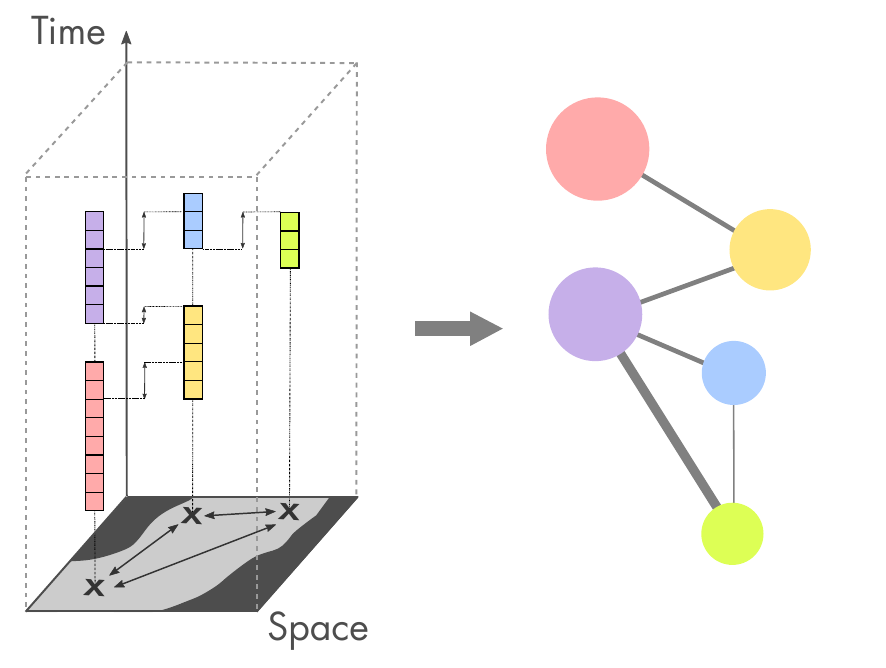}
	\caption{\textbf{Illustration of the methodology used to extract resting event networks from spatio-temporal trajectories.} Resting events are characterized by their duration, day of occurrence and location. A link is created between two events if they overlap in time on a daily basis. The weight of a link corresponds to the distance between events' location.  \label{Fig1}}
\end{figure*}

\section*{Introduction}

New technological developments of digital tracking systems contribute to the production of an ever-growing volume of high resolution animal movement data. This new source of knowledge is crucial to better understand and visualize animal movements at different scales. It can thus provide insights to develop adequate conservation planning strategies that are flexible in space and time \citep{Urbano2010,Allen2016}. However, as it has arisen recently in many disciplines, dealing with large amount of data has brought to light new problems regarding the extraction of meaningful information from huge data sets \citep{Hampton2013,Louail2015}. Handling the spatio-temporal nature of this information is one of those. 

Animal movement has long been observed and modeled through the lens of diffusive processes \citep{Turchin1998} and foraging theories \citep{Stephens1987} strongly focusing on the characteristic of the spatio-temporal trajectories such as speed or turning angles. As mentioned in \citep{Wittemyer2017}, with our new abilities to collect high resolution spatio-temporal data over long periods of time, we can more and more concentrate our research on the analysis of individual movements. We can make an analogy here between animal and human movements \citep{Meekan2017}. The tendency for human individuals for revisiting locations \citep{Gonzalez2008,Song2010} and their interactions with different types of environment according to the time of day \citep{Lenormand2015} can also be investigated in animal movement \citep{Polansky2015}. This is to some extent similar to the concept of spatial memory in animal movement (i.e. the ability to relocate to previously visited places) \citep{Fagan2013}. One can for instance focus on the habitat uses and on potential regularities in animal spatial behaviors. Some examples include the identification of repeatedly visited home range areas \citep{Benhamou2012}, the space-time characterization of springbok movement \citep{Lyons2013}, the cougars' changes in movement characteristics over time \citep{Ironside2017} and the repeated use of specific rest locations by female elephants \citep{Wittemyer2017}. However, they usually focus on long-term mobility behavior, and therefore the presence of daily spatio-temporal patterns is rarely investigated.

In this work, we are interested in identifying daily mobility patterns in fish spatio-temporal trajectories. Our main objective is to investigate the regularities characterizing the fish behaviors in space and time on a daily basis. We want to understand whether fish tend to observe the same behaviors at the same time and place each day. For this purpose, we draw upon the recent advances in individual human mobility patterns modeling and analysis \citep{Barbosa2018}, and more particularly the concept of daily motifs \citep{Schneider2013} adapted to animal movements. We rely on network-based tools \citep{Barthelemy2011,Barabasi2016} that have been widely used this past ten years in ecology in general \citep{Bascompte2007} and movement ecology in particular \citep{Jacoby2012,Jacoby2016,Bastille2018}. More precisely, we rely in this study on the concept of event network \citep{Wilson2016} that represents a powerful tool to extract a coarse-grained signature of spatio-temporal trajectories. We are interested in the connections between resting events, defined as the presence of an individual in a particular location during a time windows higher than a predetermined threshold. Hence, the nodes of the considered networks are defined in space and time and connected according to their spatio-temporal proximity. The analysis of these networks will enable us to uncover daily fish resting patterns.

The next section describes in details the proposed approach. The guiding idea is that resting event networks can be extracted from spatio-temporal (Figure \ref{Fig1}). These networks are then analyzed and compared with a null model preserving the observed events spatio-temporal characteristics to ensure that the patterns identified are not due to random configurations. Network science offers a wide variety of tools and metrics to explore systematically the event network structure. We are particularly interested in the event networks' topological structure and its degree distribution investigating the relationship between the connection of events in time and their spatial proximity. Sets of highly connected events and the movements between them are also considered in order to identify statistically prevalent network communities and network motifs. 

We apply the method to analyze the daily mobility structure of different fish species in the Rh\^one River located in France. Although the temporal structure of the resting event network is mainly driven by the distribution of events duration and their day of occurrence, we show that it exists a spatial proximity between event occurring at similar hours but on different days. Finally, the method allows to capture daily mobility motifs in the global event network structure that are not reproduced by the null model.

\section*{Methods}

\subsection*{Data}

The data set used for our analysis contains information about fish positions recorded with acoustic telemetry techniques between July and September 2009 in the Rh\^one River (France). In acoustic telemetry, the fish are tagged with acoustic transmitters that are then detected by receiver stations deployed in their natural environment. These data were collected as part of a research project conducted in a 1.8 km long and 140 m wide river segment. The purpose of the project was to track the movements of 94 fish captured in June 2009 in the river segment. For more details about the experiment see \citep{Berge2012,Capra2017,Lamonica2020}. In favorable areas, fish position can be received every 3 seconds. However, the signal can be subject to discontinuity in certain area of the river segment. Moreover, the presence of tagged fish in the study area can be very irregular and highly dependent on the fish individuals and the fish species. To assess the quality of the individual fish data, we segment each day of observation into 288 5-minute periods and compute the fraction $\gamma$ of periods during which the position of the fish was recorded at least once. Based on this metric, we selected ten fish among the most frequently localized individuals that belong to three species: four barbels (\textit{barbus barbus}), two catfishes (\textit{ictalurus melas}) and four chubs (\textit{squalius cephalus}). For each fish individual, we selected the ten days exhibiting the highest $\gamma$ values. On average we detected the presence of the fish in the study area $85\%$ of the day, with a minimum presence of $60\%$ and a maximum of $100\%$. More details regarding the fish and day selection processes are available in mentary Information (Figure S1 and Figure S2).

\subsection*{Daily spatio-temporal trajectory}

Fish trajectories are characterized by a sequence of visited locations. To build these sequences, both time and space need to be discretized. Each day is segmented into 288 5-minute periods and the river segment is divided into a regular grid composed of square cells of lateral size 20 meters. Each 5-minute period  is assigned a location (i.e. a grid cell) if a position was recorded in that time interval. If no position is recorded during a time period, we assign it an unknown location. If the presence of a fish is detected into several grid cells in a given 5-minute period, we choose the cell with the highest number of records. In the event of a tie, one of them is drawn at random. Nevertheless, in most of the cases, fish individuals spend most of their time in one location during a time interval (Figure S3a in Appendix). At the end of the process, we obtain 100 daily spatio-temporal trajectories (ten days for each of the ten selected fish). A daily trajectory is represented by a spatio-temporal sequence $S=\{X_1,...,X_T\}$ of locations at which a fish was observed at each consecutive 5-minute interval ($T=288$). It is important to note that some of these locations are unknown. However, the periods during which the presence of a fish is not detected in the study area during the selected days represents on average less than $15\%$ of the time. Moreover, consecutive time periods with unknown location last generally less than fifteen minutes (Figure S3b in Appendix).

\subsection*{Resting event networks}

The daily spatio-temporal trajectories defined in the previous section can be decomposed in a succession of events devoting to different fish \enquote{activities}. An event $e$ is a sub-sequence $S_e \subseteq S$ of consecutive locations. It is characterized by a starting time period $t_e$ and a duration $\Delta_e$. In this work, we consider that a resting event $r$ occurs when a fish rests in the same location during at least $\lambda$ consecutive time periods ($\Delta_e \geq \lambda$). We assume that unknown locations are always associated with non resting event whatever their duration. We only consider resting event starting and ending during the day (i.e. $t_e > 1 $ and $t_e + \Delta_e - 1 < 288 $).      

For each fish, we obtain a collection of resting events $R$ representing every resting events identified among the ten daily spatio-temporal trajectories. Whether an event belongs or not to $R$ depends on the threshold $\lambda$. Indeed, if $\lambda=1$ all the events are considered as resting events, and inversely, if $\lambda > 288$ the entire trajectory will be consider as a non resting event. We may assume that the chosen value will depend on the type of animal but, in our case, the value $\lambda=3$ (15 minutes) seems to be a good compromise allowing us to preserve a reasonable number of resting events per day (between 6 and 28) while minimizing the variability across daily spatio-temporal trajectories (see Figure S4 in Appendix for more details).

Now that the nodes of the resting event networks are formally defined, we need to connect them according to their similarities from both the spatial and temporal point of views. To this end, we propose two similarity metrics, $\delta_t$ and $\delta_s$, to link the events according to their spatio-temporal proximity. $\delta_t$ computes the number of time periods shared by two events while $\delta_s$ measures the spatial proximity between two events $e$ and $e'$ based on the distance $d_{ee'}$ between the event locations (Equation \ref{deltas}). To be more specific, the distance $d_{ee'}$ is equal to the euclidean distance between the centroids of the cells where the events $e$ and $e'$ occurred (expressed in meters).
\begin{equation}
\delta_s(e,e')=\frac{1}{1+d_{ee'}}
\label{deltas}
\end{equation}
In this work, we decided to focus on the temporal proximity to build the topological structure of the networks and on the spatial proximity to define the intensity of interactions between events. More specifically, a link is created between two events $e \neq e'$ if $\delta_t(e,e')>0$ and the weight of a link between them is equal to $\delta_s(e,e')$. The creation of a link thus implies that two events share at least one time period.  Since two events occurring at the same day do not overlap in time, it therefore follows that there is no link between events occurring at the same day. This is an important characteristic of the resting event networks that we propose in this study. At the end of the process, we obtain one weighted undirected spatio-temporal resting event network per fish. 

\subsection*{Null model}

To properly characterize the event networks and identify potential daily mobility patterns in fish trajectories we first need to define a null model (NM). Null model analysis are really useful to identify non-random patterns. In our case we need to generate random event networks preserving the observed events spatio-temporal characteristics: the number of events, the events duration and day of occurrence, and the global spatial distribution of events. The topology of the resting event network introduced in the previous section is strongly constrained in time. Indeed, the probability $\mathbb{P}(\delta_t(e,e')>0)$ of connecting two events in a random situation is highly dependent of the events' duration and whether they occurred on the same day or not. We can however take these temporal constraints into account by generating random networks' topology in which, for a given day, starting events time are drawn at random along the day. In other words, we reshuffled, for each fish, the starting time of every events of the resting event network while preserving the day of occurrence of the events and their duration. Regarding the spatial component of the network (i.e. link weights), we generated random weights $\delta_s(e,e')$ by reshuffling the resting events' location, thus preserving the spatial distribution of events locations over the ten days of observation. Using this approach we generate $100$ random event networks for each fish.

\begin{figure*}
	\centering 
	\includegraphics[width=\linewidth]{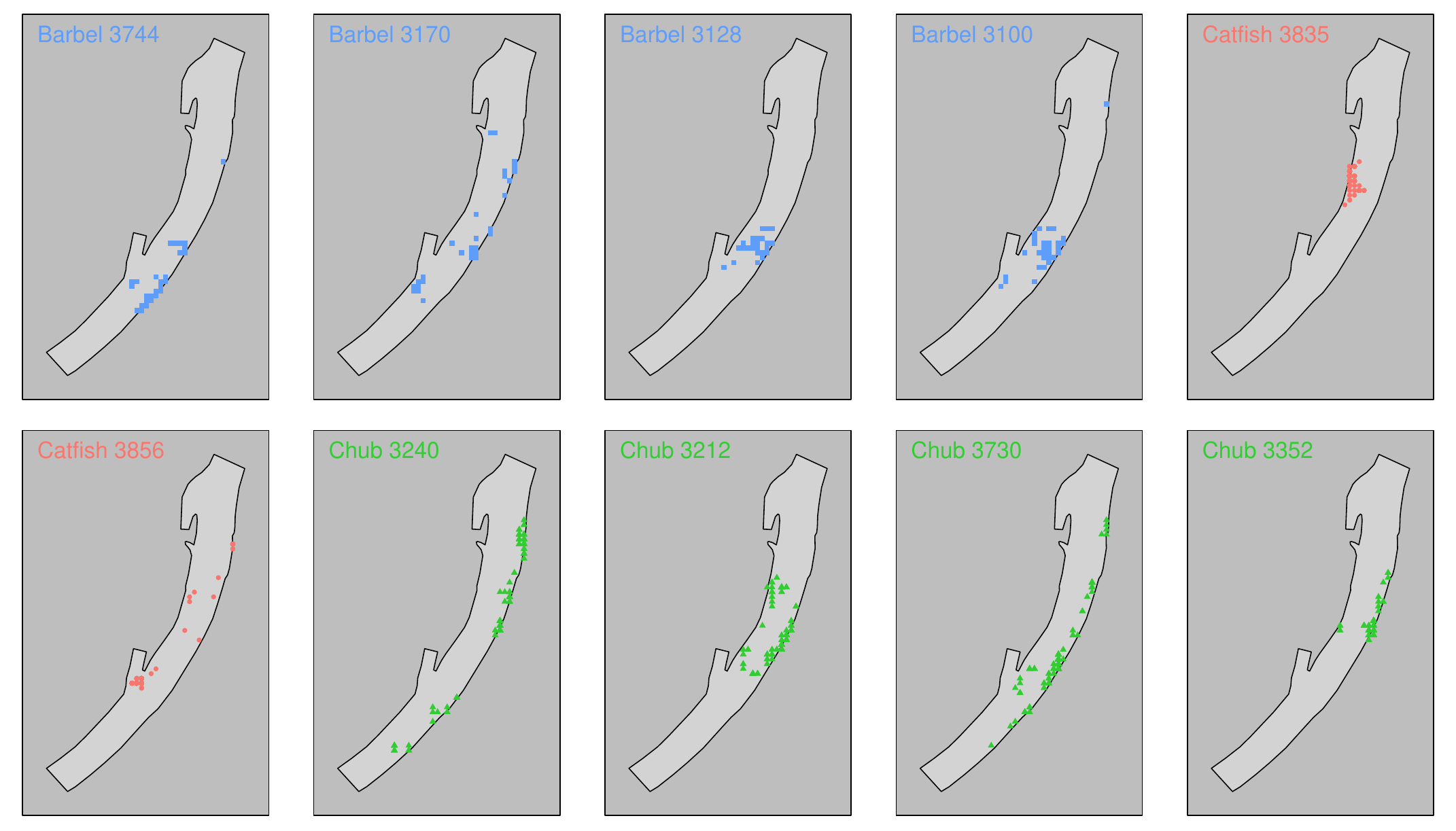}
	\caption{\textbf{Resting event's spatial distribution.} The blue squares represent the barbels' resting event locations, the red squares stand for the catfish and the green circles for the chubs. \label{Fig2}}
\end{figure*}

\subsection*{Network measures}

\noindent \textit{Degree.} Networks topology can be quantitatively described by a wide variety of measures. The most important of them is probably the node degree. The degree of a node is the number of connections that links it to the rest of the network. To evaluate to what extent the degree distributions are characteristic of the event networks structure, we will compare these distributions to the ones returned by the null models.\\ 

\noindent \textit{Dilatation index.} Resting event networks are also spatial networks. To characterize the spatial component of event networks we introduce the Dilatation Index (DI) defined as the average pairwise euclidean distance between connected events. This metric is expressed in meters and defined as follows,
\begin{equation}
DI=\frac{1}{|A|}\sum_{(e,e') \in A} d_{ee'}
\label{DI}
\end{equation}
where $A=\{(e,e') \in R \times R \,\, | \,\, e \neq e' \,\land\, \delta_t(e,e')>0\}$ represents the set of pairs of connected events. As defined above, $d_{ee'}$ is equal to the euclidean distance between the centroids of the cells where the events $e$ and $e'$ occurred (expressed in meters). In order to contrast the results, two other dilatation indices are also considered, $DI_{tot}$ defined as the average pairwise euclidean distance between all the events (i.e. $A=\{(e,e') \in R \times R \,\, | \,\, e \neq e'\}$), and, $DI_{NM}$ defined as the average pairwise euclidean distance between connected events generated with the null model described above.\\

\noindent \textit{Network community structure.} Community structure is an important network feature, revealing both the network internal organization and similarity patterns among its individual elements. In this study we used the Order Statistics Local Optimization Method (OSLOM) algorithm proposed in \citep{Lancichinetti2011} that detects statistically significant network community with respect to a global null model (i.e. random graph without community structure). This algorithm is non-parametric in the sense that it returns the optimal statistically significant partition without defining the number of communities \textit{a priori}. More details about OSLOM are available in Appendix. In our case, the purpose is to identify spatio-temporal communities clustering events exhibiting significant temporal and spatial proximity.\\

\begin{figure*}
	\centering 
	\includegraphics[width=\linewidth]{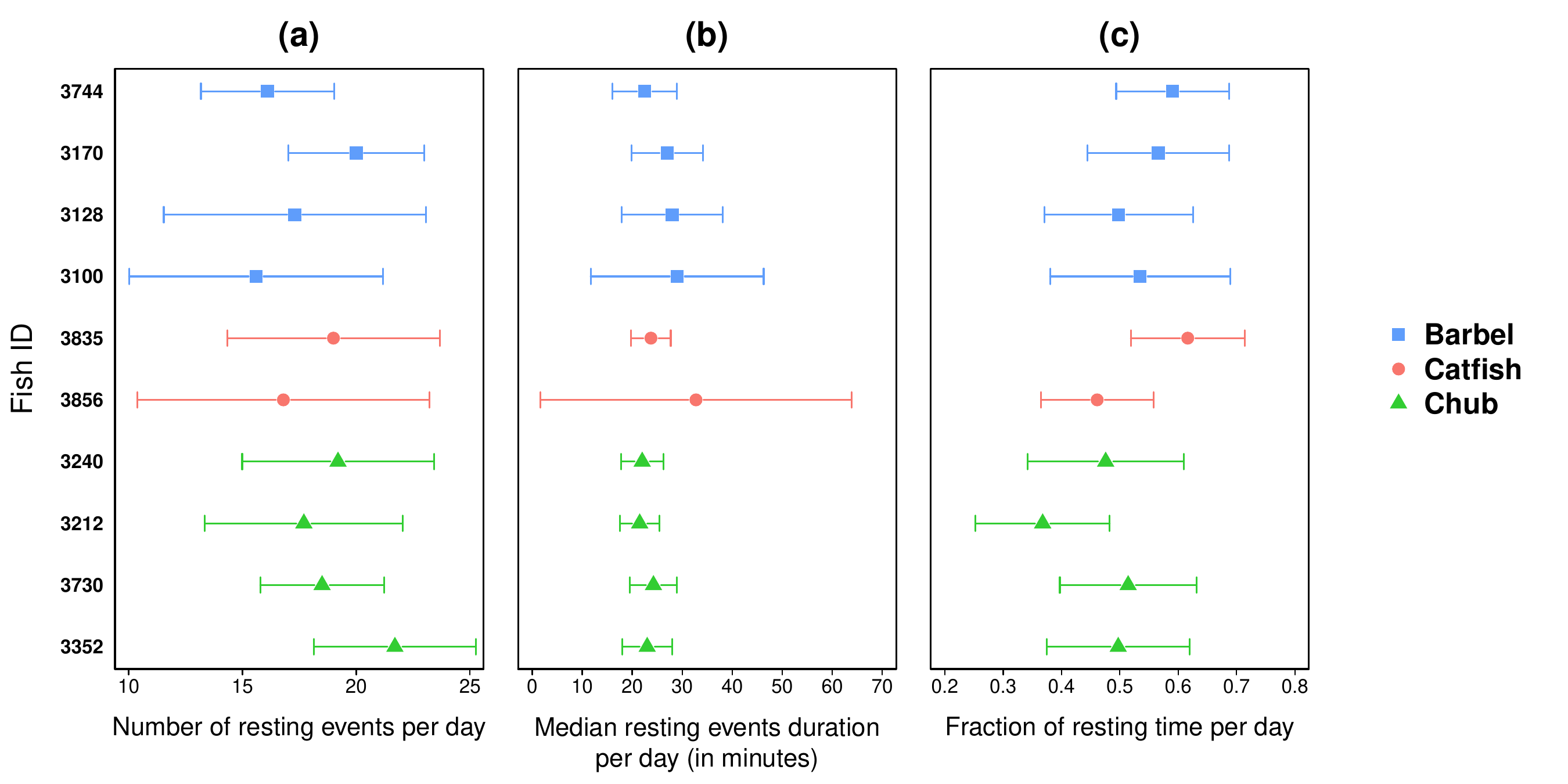}
	\caption{\textbf{Characteristics of the resting events according to the fish and species.} (a) Average number of resting events per day. (b) Average event median duration per day in minutes. (c) Fraction of resting time per day. The values have been averaged over the ten days of observation for each fish. The error bars represent one standard deviation. \label{Fig3}}
\end{figure*}

\noindent \textit{Network motifs.} An interesting local network property is recurrent patterns, repeating themselves in a network, and usually called network motifs. In this study, a motif is defined as a displacement between spatio-temporal communities. To be more specific, two consecutive resting events occurring the same day can be interpreted as a spatio-temporal displacement between the community to which the first event belong to the community of the second event. We call this displacement a motif characterized by a movement between a community of origin and a community of destination that can be identical. A motif can be seen as an ordered pair of communities. Hence, for every fish and day of observation we can extract a list of daily motifs. Similarly to the method used in \citep{Clemente2018}, the S{\o}rensen index \citep{Sorensen1948} is used to define matrices of similarity between lists of daily motifs. This index varies from $0$, when no agreement is found, to $1$, when the two lists are identical. For each fish, we obtain 45 comparisons, each of them assessing the motifs similarity between two days of observation that can be used to investigate daily mobility patterns. More detail about the method used to compute the similarity between daily motifs are available in Appendix.

\section*{Results}

\subsection*{Resting event networks}

\noindent \textit{Resting events.} In order to get a preliminary grasp of the data we plot the resting event's spatial distribution in Figure \ref{Fig2}. We observe that the resting event locations are more or less dispersed according to the fish individual. It seems however that there is no significant differences among species. We also plot several resting event characteristics in Figure \ref{Fig3}. Despite some particularities according to the species, the selected fish shows globally similar event features. The average number of resting events per day, displayed in Figure \ref{Fig3}a, lies between 15 and 20 with a standard deviation of 3 days. Regarding the duration of these events, fish tend to rest half of the day in average (Figure \ref{Fig3}c) with a median resting event duration around 20 - 30 minutes (Figure \ref{Fig3}b). Although the difference is not significant some differences among species can be observed. Chubs tend to have a higher number of resting events but with a lower fraction of resting event duration and resting time than the two others species. It is also interesting to note that their resting event characteristics are more stable in time (i.e. day) particularly regarding the resting event median duration.\\

{\renewcommand{\arraystretch}{1.2}	
	\begin{table*}[t]
		\caption{\textbf{Statistical properties of the resting event networks.} Number of events (\#Nodes), number of links (\#Links), average degree (Degree) and dilatation indices (expressed in meters). All the metrics based on the null model (NM) have been averaged over 100 replications. The associated standard deviations are available in Appendix.}
		\label{Tab1}
		
		\vspace*{0.5cm}
		\hspace*{-1cm}
		%\begin{center}
		\begin{tabular}{cccccccccc}			
			\hline 
			\textbf{Fish ID} & \textbf{Species} & \textbf{\#Nodes} & \textbf{\#Links} & \textbf{\#Links (NM)} & \textbf{Degree} & \textbf{Degree (NM)} & \textbf{$DI$} & \textbf{$DI_{tot}$} & \textbf{$DI_{NM}$}\\
			\hline
			
			3744 & Barbel & 161 & 675 & 917.75 & 8.39 & 11.42 & 211.50 & 241.75 & 238.96 \\ 
			3170 & Barbel & 199 & 1049 & 1071.55 & 10.54 & 10.72 & 584.66 & 633.38 & 633.09 \\ 
			3128 & Barbel & 173 & 846 & 812.62 & 9.78 & 9.41 & 171.02 & 179.30 & 180.14 \\ 
			3100 & Barbel & 156 & 733 & 798.18 & 9.40 & 10.25 & 300.96 & 384.80 & 381.70 \\ 
			3835 & Catfish & 190 & 944 & 1116.14 & 9.94 & 11.76 & 157.94 & 169.11 & 167.06 \\ 
			3856 & Catfish & 168 & 668 & 728.65 & 7.95 & 8.70 & 206.75 & 135.76 & 131.74 \\ 
			3240 & Chub & 192 & 772 & 848.21 & 8.04 & 8.85 & 444.92 & 460.90 & 462.58 \\ 
			3212 & Chub & 176 & 507 & 596.46 & 5.76 & 6.78 & 292.03 & 298.11 & 297.49 \\ 
			3730 & Chub & 184 & 852 & 891.91 & 9.26 & 9.65 & 351.87 & 375.32 & 375.84 \\ 
			3352 & Chub & 216 & 942 & 1001.87 & 8.72 & 9.24 & 157.95 & 171.40 & 171.71 \\ 	
			
			\hline			
		\end{tabular}
		%\end{center}
	\end{table*}
}

\noindent \textit{Event network topology.} We now want to identify potential temporal patterns by comparing the observed event network topology with the one returned by the null model introduced in the methods section. Some basic network properties are gathered in Table \ref{Tab1}. The random network has more connections than the original one leading to a slightly higher average degree. The event network degree distribution is an important feature that allows for the identification of temporal patterns. However, the degree distribution alone is not very informative since a network constrained in time will naturally tend to exhibit a heavy tail distribution. A comparison with the degree distribution obtained with the null model is therefore crucial to identify any particular network topology. Figure \ref{Fig4} displays an example of such comparison for a barbel (see Figure S5 in Appendix for all fish). Although the observed and randomized degree distributions (top and left insets) are similar, the observed degree of a specific event can be very different from the one returned by the null model. This deviation from the random situation is a good indicator of the presence of patterns in resting events temporal distribution. We observe that the degree of events with a very low observed degree increased systematically in random situation. It is important to note here that there is a clear relationship between the event duration and its degree. Indeed, the higher the duration of an event, the higher the probability for this event to be connected to other events (see Figure S6 for more details). Short resting events are therefore less connected than they should be. Conversely, the highly connected events of some fish (Figure S5) are less connected in the observed network than in random conditions. It means that events that should be connected according to their long duration and day of occurrence are not connected. However, this is not the case for most of the fish whose temporal event network structure does not deviate substantially from the one returned by the null model (see Figure S5).

\begin{figure}[!ht]
	\centering 
	\includegraphics[width=\linewidth]{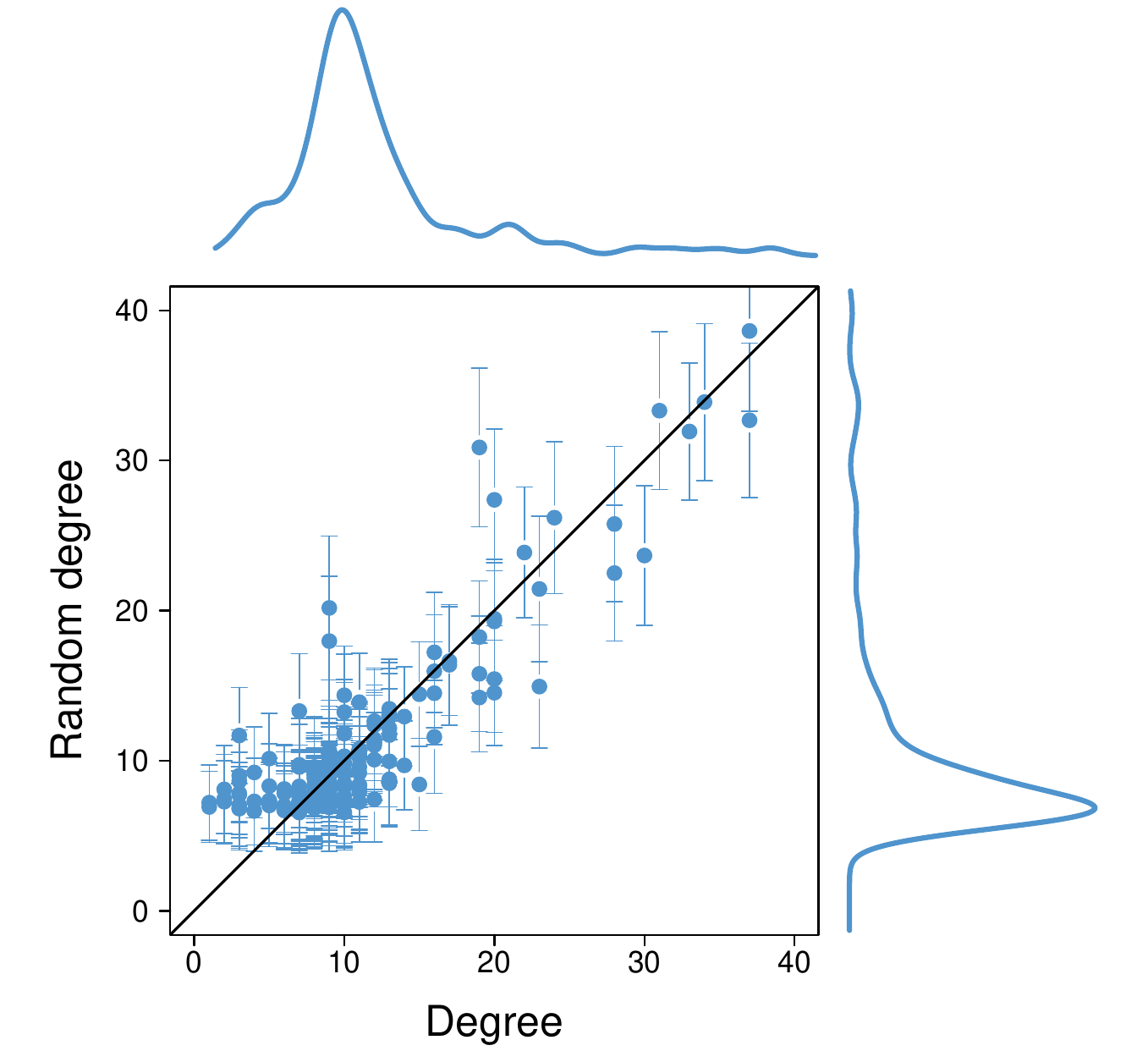}
	\caption{\textbf{Comparison between observed and average random degree distribution.}  Each point represents an event with the observed average degree on the x-axis and the average degree obtained with the null model on the y-axis averaged over 100 replications (the error bars represent one standard deviation). The insets show the marginal probability density distribution. We choose a representative example with the Barbel 3170. Similar plots for all fish are available in Figure S5 in Appendix.} \label{Fig4}
\end{figure} 

To conclude, we observe differences between observed and random network topologies. Nevertheless, the presence of patterns in the temporal structure of the fish event networks remains unclear. For most fish it seems indeed mostly driven by the distribution of events duration and their day of occurrence than specific temporal patterns.\\

\begin{figure*}
	\centering 
	\includegraphics[width=13.5cm]{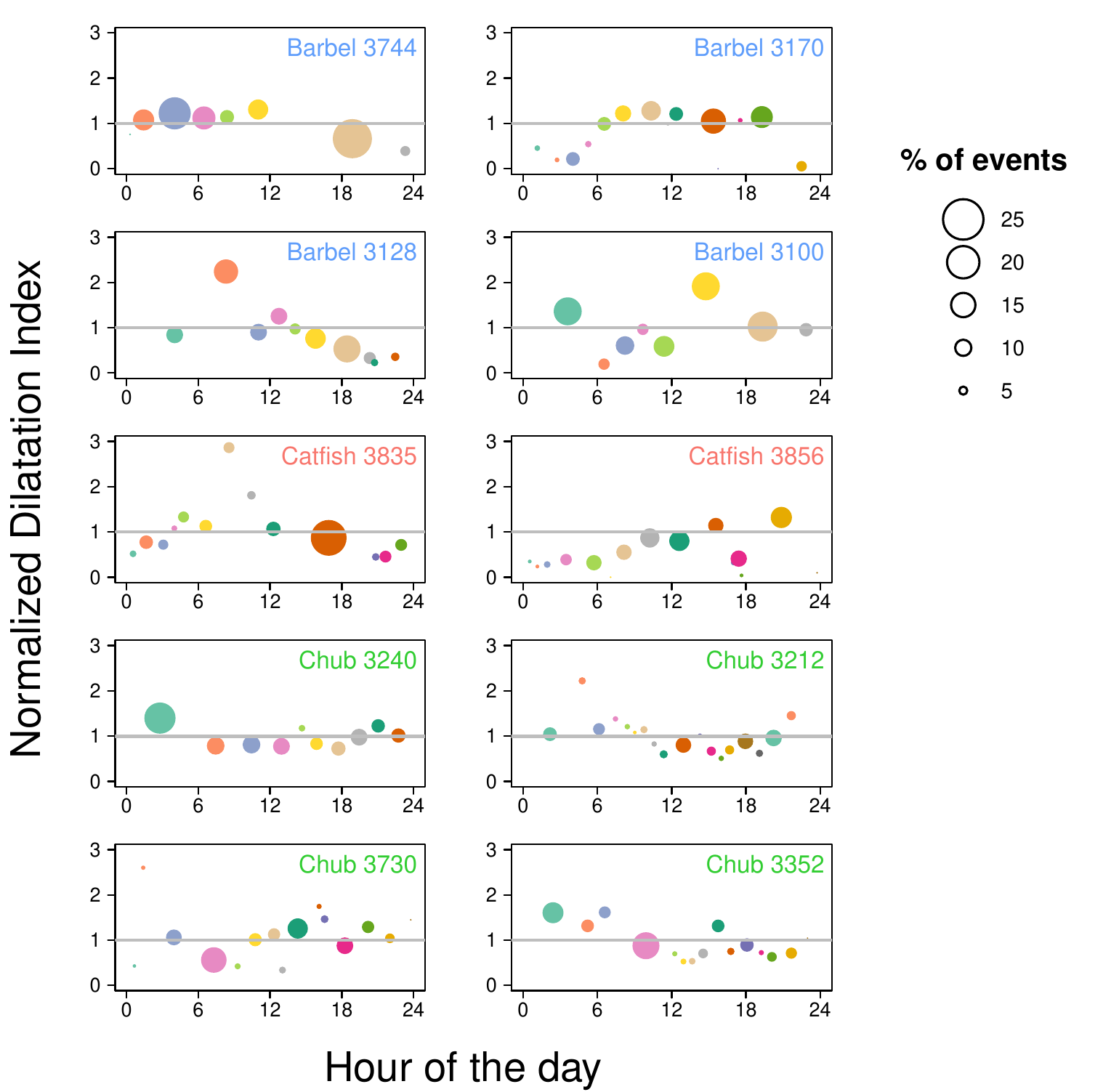}
	\caption{\textbf{Analysis of the event networks' community structure.} The plots display the communities characteristics for every fish. Each point represents a community with the average resting time on the x-axis and the community dilatation index on the y-axis. The community dilatation index is normalized by the dilatation index $DI$. The size of the dots is proportional to the fraction of events. \label{Fig5}}
\end{figure*}

\noindent \textit{Spatio-temporal structure.} To investigate the relationship between space and time in the resting event distribution we compute the dilatation index based on fish event locations for different network configurations. All the metrics based on the null model (NM) have been averaged over 100 replications. Table \ref{Tab1} presents the results obtained for each fish. We observe that except for the second catfish the dilatation index measured in the observed network ($DI$) is lower than the one measured in the null model ($DI_{NM}$). The latter is actually equivalent to the dilatation index  $DI_{tot}$ defined as the average distance between all the events not only the connected ones. This observation suggests that there are spatio-temporal patterns hidden in the fish resting event networks analyzed in this study. Implying that a temporal proximity between resting events leads to a spatial proximity between events' location. 

\subsection*{Event network community analysis}

In order to go further in the analysis of fish resting events spatio-temporal structure we perform a network community analysis for each of the ten selected fish. We first rely on the number of communities to assess the community structure obtained with the OSLOM algorithm. We observe in Table \ref{Tab2} that resting events can be globally clustered in a dozen of spatio-temporal communities. Note that this number can vary by a factor of two from one fish to another. Chubs tend to have more communities than the other fish, probably due to the fact that they have more and shorter resting events than the two others species. Figure \ref{Fig5} shows a representation of the spatio-temporal distribution of communities according to their size (i.e. number of events). The temporal dimension is presented on the x-axis with the average time (hour of the day) at which the events occurred. The spatial dimension is presented on the y-axis with the community dilatation index between connected events belonging to a community normalized by the \enquote{global} dilatation index ($DI$ in Table \ref{Tab1}). We observe different size of communities, the biggest community contains in average 17\% of the events, but there is no evidence of existence of a relationship between the community size and its average time of occurrence. 

{\renewcommand{\arraystretch}{1.2}
	\begin{table}[!ht]
		\caption{\textbf{Number of resting event network communities.} The metric based on the null model (NM) have been averaged over 100 replications. The associated standard deviations are available in Appendix.}
		\label{Tab2}
		\begin{center}
			\begin{tabular}{cccc}			
				\hline 
				\textbf{Fish ID} & \textbf{Species} & \textbf{\#Com} & \textbf{\#Com (NM)}\\
				\hline
				
				3744 & Barbel &   8 & 7.41\\ 
				3170 & Barbel &  15 & 12.51\\ 
				3128 & Barbel &  10 & 13.12\\ 
				3100 & Barbel &   8 & 10.23\\ 
				3835 & Catfish &  13 & 9.98\\ 
				3856 & Catfish &  15 & 12.17\\ 
				3240 & Chub &  10 & 11.03\\ 
				3212 & Chub &  19 & 14.66\\ 
				3730 & Chub &  16 & 9.96\\ 
				3352 & Chub &  16 & 18.86\\  	
				
				\hline			
			\end{tabular}
		\end{center}
	\end{table}
}

\begin{figure*}
	\centering 
	\includegraphics[width=15cm]{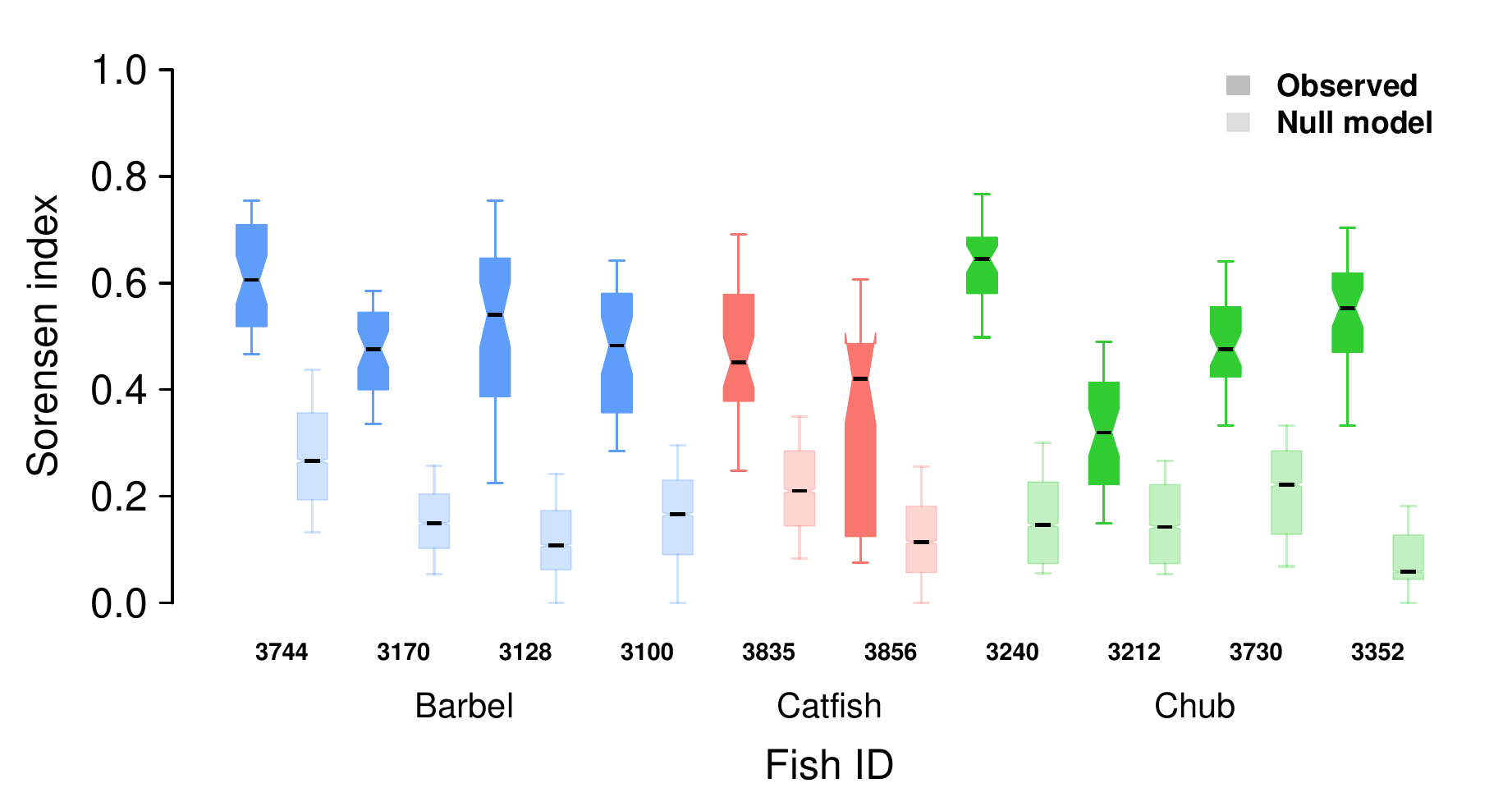}
	\caption{\textbf{Similarity between fish daily motifs.} Boxplot of S{\o}rensen index between fish daily motifs according to the fish and fish species. Results obtained with the null model are displayed in a brighter version of the baseline fish species color. The boxplot is composed of 45 comparisons for the observed motifs and 45*100 comparisons for the motifs obtained with the null model. Each boxplot is composed of the first decile, the lower hinge, the median, the upper hinge and the last decile. \label{Fig6}}
\end{figure*} 

We have shown in the previous section that a relationship between the temporal proximity of events and their spatial proximity exists. We observe in Figure \ref{Fig5} that this spatial proximity between connected events varies along the day with a community dilatation index more or less close to the global one (grey line) according to the hour of the day. Some communities exhibit a high dilatation index, up to three times $DI$, while others show a low dilatation index, sometimes close to 0. These deviations from the average may suggest different types of fish daily behaviors related to the heterogeneity of visited places according to the hour of the day.  

In order to assess the significance of these results, we analyze the community structure, community size (Table \ref{Tab2}) and their spatio-temporal distribution (Figure S7), obtained with the null model. As in Figure \ref{Fig5}, Figure S7 represents the spatio-temporal distribution of communities but for one realization of the null model. In this case the dilatation index between connected events belonging to the communities is normalized by the \enquote{global} dilatation index obtained with the null model ($DI_{NM}$ in Table \ref{Tab1}). The differences between observed and null communities in terms of number and size is not striking. However, the difference between global and community dilatation indices is lower for the resting network obtained with the null model that the observed resting event networks. We have already shown that the dilatation index between connected events is significantly higher in random situation than in the observed one, but we also observed a temporal variation of the community dilatation index, not reproduced by the null model. 

It is however not clear whether or not these spatio-temporal patterns correspond to regularities due to the presence of fish daily motifs.

\subsection*{Daily motifs}

With these event communities, we can now assess the similarity between fish daily motifs. As described in the methods section, each day of observation of a fish can be represented by a list of network motifs defined as a intra- or inter- community displacement. We then calculate the S{\o}rensen index between the ten lists of daily motifs for each fish. Figure \ref{Fig6} shows notched boxplots of the S{\o}rensen index obtained with the observed daily motifs and the ones returned by the null model. First, we observe that the similarity between daily motifs is globally high, with a median percentage of motifs in common ranging between 30 and 60 percent. It is worth noting that some days are more similar than others with a high variability around the median value. It seems however that there is no significant differences among species.  

It is really interesting to note that the similarity between daily motifs is always significantly higher in the observed than the random situation (between 2 and 3 times higher). Therefore, the daily mobility motifs identified here are not due to random configuration, they are the sign of spatio-temporal regularities in fish daily mobility behaviors.

\section*{Discussion}

Being able to develop new statistical tools and methods to extract meaningful information from large data sets is crucial to enhance our comprehension of ecological systems. In this study, we contribute to this end by proposing a general method based on the concept of resting event network to analyze animal daily mobility patterns. We successfully applied this method on several fish species in a large hydropeaking river in France. In particular, we showed that, despite some particularities according to the species, resting events characteristics are remarkably stable among fish individuals. We also found that, despite a few exceptions, the temporal dimension of resting event structure is mainly driven by the distribution of events duration and their day of occurrence. However, the spatial proximity between events temporally connected is higher in the observed events than the ones generated with the null model. This finding has been confirmed with the network community analysis performed in this study, showing that the community structure in terms of number and size is very similar to the ones return by the null model but the presence of temporal variation of the spatial component of the communities is not reproduced by the null model. Finally, we extracted daily motifs and demonstrated the presence of significant regularities in daily fish mobility.

The example chosen to illustrate the methodology is based on a local data set and on a small sample of individuals. It would be interesting to apply the proposed approach to other animals such as big terrestrial and marine mammalians for example. Nevertheless, focusing on fish daily mobility pushed us to incorporate a null model in the analysis enabling us to put aside patterns due to spatio-temporal constraints but also to highlight non random regularities.

To conclude, given the importance of animal resting behaviors in conservation planning strategies, the future application and adaptation of the proposed methodology are numerous. Moreover, as it is often the case with network-based tools, we believe that a key feature of the proposed approach resides in its generic nature since it can be applied to any type of individual spatio-temporal trajectories.

\section*{Acknowledgments}

This work was supported by a grant from the French National Research Agency (project NetCost, ANR-17-CE03-0003 grant). We wish to thank Dino Ienco and Riccardo Gallotti for useful discussions. We thank the Agence de l'Eau Rh{\^o}ne-M{\'e}diterran{\'e}e-Corse, Electricit{\'e} de France (EDF-DTG), the European Union/FEDER and the Aquitaine region for their financial support. We gratefully acknowledge the HTI engineers (Tracey Steig, Patrick Nealson, David Ouellette and Samuel Johnston) for their valuable assistance in the processing of acoustic data. Finally, we thank Pascal Roger, Raphael Mons, Fr{\'e}d{\'e}rique Bau, Olivier Croze, C{\'e}dric Giroud (professional fisherman) and the numerous people who contributed to field work.

\bibliographystyle{unsrt}
\bibliography{EventNetwork}

\onecolumngrid
\vspace*{2cm}
\newpage
\onecolumngrid

\makeatletter
\renewcommand{\fnum@figure}{\sf\textbf{\figurename~\textbf{S}\textbf{\thefigure}}}
\renewcommand{\fnum@table}{\sf\textbf{\tablename~\textbf{S}\textbf{\thetable}}}
\makeatother

\setcounter{figure}{0}
\setcounter{table}{0}
\setcounter{equation}{0}

\newpage
\clearpage
\newpage
\section*{Appendix}

\section*{The OSLOM algorithm}

In this study we used the Order Statistics Local Optimization Method (OSLOM) \cite{Lancichinetti2011} to identify the communities in the resting events networks . OSLOM uses an iterative process to detect statistically significant communities with respect to a global null model (i.e. random graph without community structure). The main characteristic of OSLOM is that it is based on a score used to quantify the statistical significance of a cluster in the network \cite{Lancichinetti2010}. The score is defined as the probability of finding the cluster in a random null model. The random null model used in OSLOM is the configuration model \cite{Molloy1995} that generates random graphs while preserving an essential property of the network: the distribution of the number of neighbors of a node (i.e. the degree distribution). Therefore, the output of OSLOM consists in a collection of clusters that are unlikely to be found in an equivalent random network with the same degree sequence. This algorithm is nonparametric in the sense that it identifies the statistically significant partition, without defining the number of communities \textit{a priori}. However, the \textit{tolerance} value that determines whether a cluster is significant or not might play an important role for the determination of the clusters found by OSLOM. The influence of this value, fixed initially, is however relevant only when the community structure of the network is not pronounced. When communities are well defined the results of OSLOM do not depend on the particular choice of \textit{tolerance} value \cite{Lancichinetti2011}. See \cite{Lancichinetti2011} for a comparison between OSLOM and other community detection algorithms.

\section*{Similarity between daily motifs}

Let us consider two lists of daily motifs $m1=[15 \rightarrow 15, 15 \rightarrow 15, 15 \rightarrow 6, 6 \rightarrow 5]$ and $m2=[15 \rightarrow 15, 15 \rightarrow 6, 6 \rightarrow 7, 7 \rightarrow 6, 6 \rightarrow 5]$. Each motif represents a displacement between communities. Displacements inside the same community are considered as valid motifs. It is also important that the same motif may appear several times in the list of daily motif. The similarity between $m_1$ and $m_2$ is defined as follows using the S{\o}rensen index,  
\begin{equation}
S=\frac{2|m_1 \cap m_2|}{|m_1|+|m_2|}
\label{S}
\end{equation}
In our example the similarity between $m_1$ and $m_2$ is equal to $S=2*3/9=2/3$.

\newpage
\section*{Supplementary Figures}

\begin{figure}[!ht]
	\centering 
	\includegraphics[width=\linewidth]{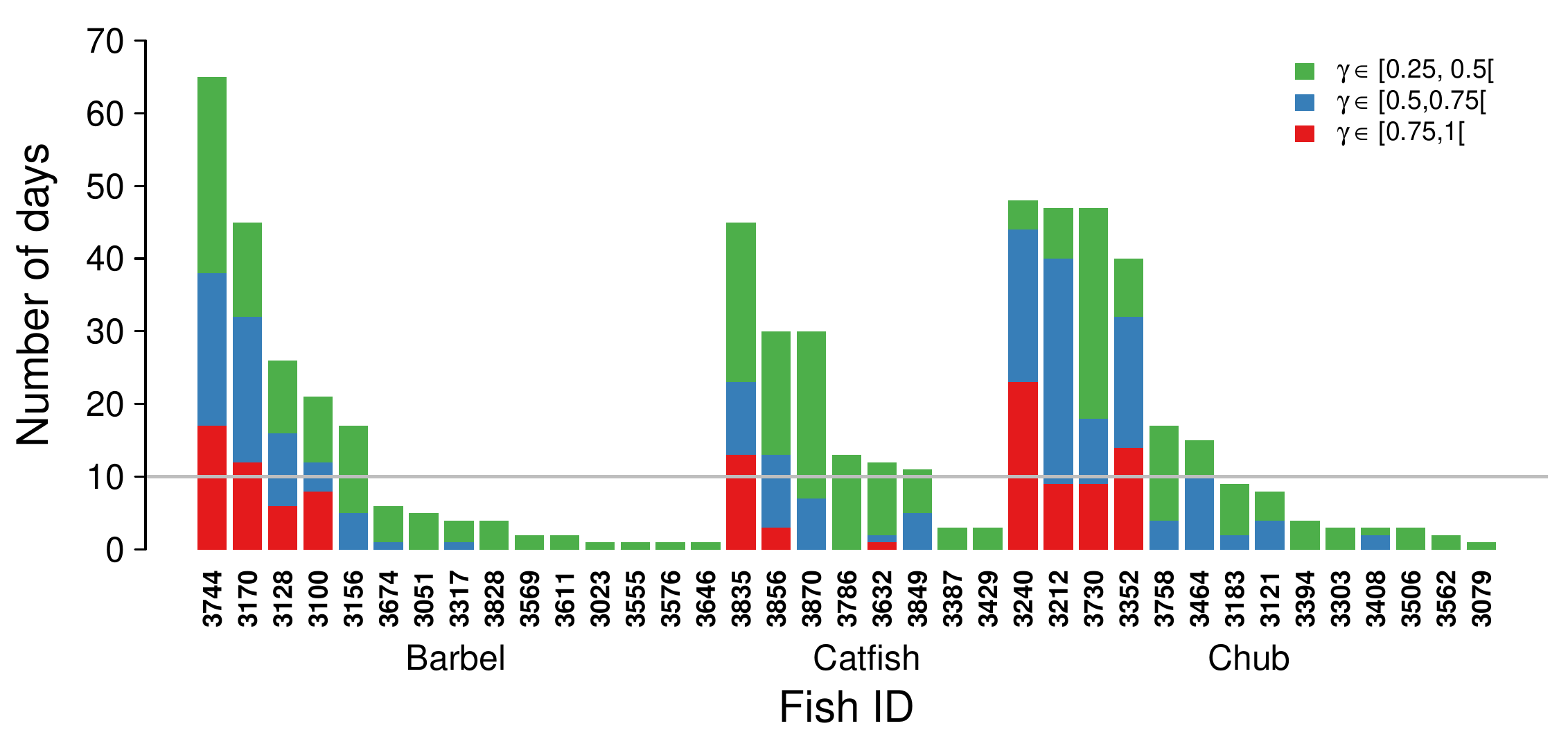}
	\caption{\textbf{Number of days of presence in the study site.} $\gamma$ represents the fraction of \emph{5-minute periods} during which the position of the fish was recorded. Three groups of values have been considered ($\gamma \in [0,0.5[$, $[0.5,0.75[$ and $[0.75,1]$). Four barbels (3744, 3170, 3128 and 3100), two catfishes (3835 and 3856) and four chubs (3240, 3212, 3730 and 3352) have been selected. All selected fish are present in the study area at least half of the day ($\gamma > 0.5$) for at least 10 days (grey line). \label{FigS1}}
\end{figure}

\begin{figure}[!ht]
	\centering 
	\includegraphics[width=14cm]{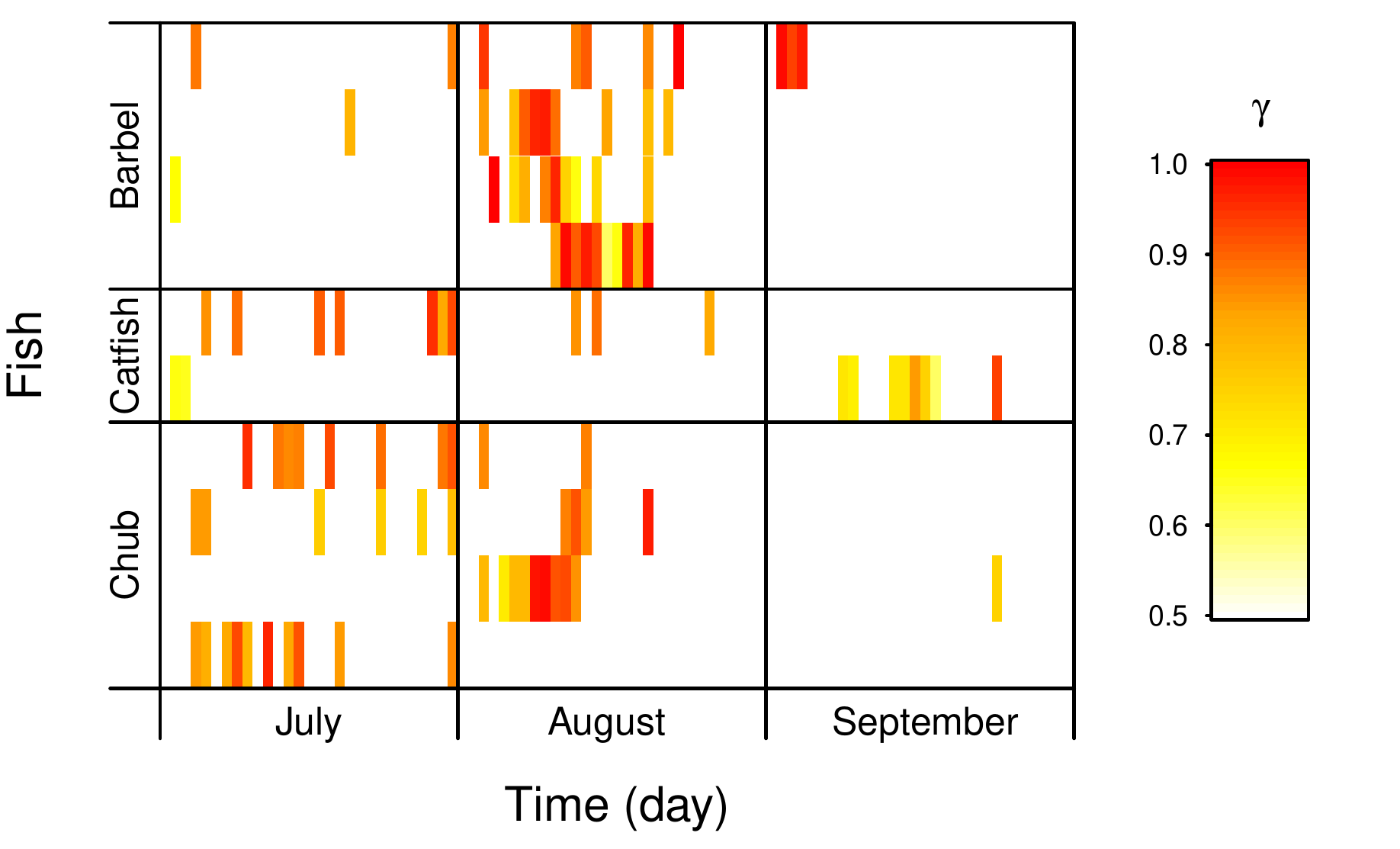}
	\caption{\textbf{The ten days with highest $\gamma$ values for each selected fish.} \label{FigS2}}
\end{figure}

\begin{figure}[!ht]
	\centering 
	\includegraphics[width=\linewidth]{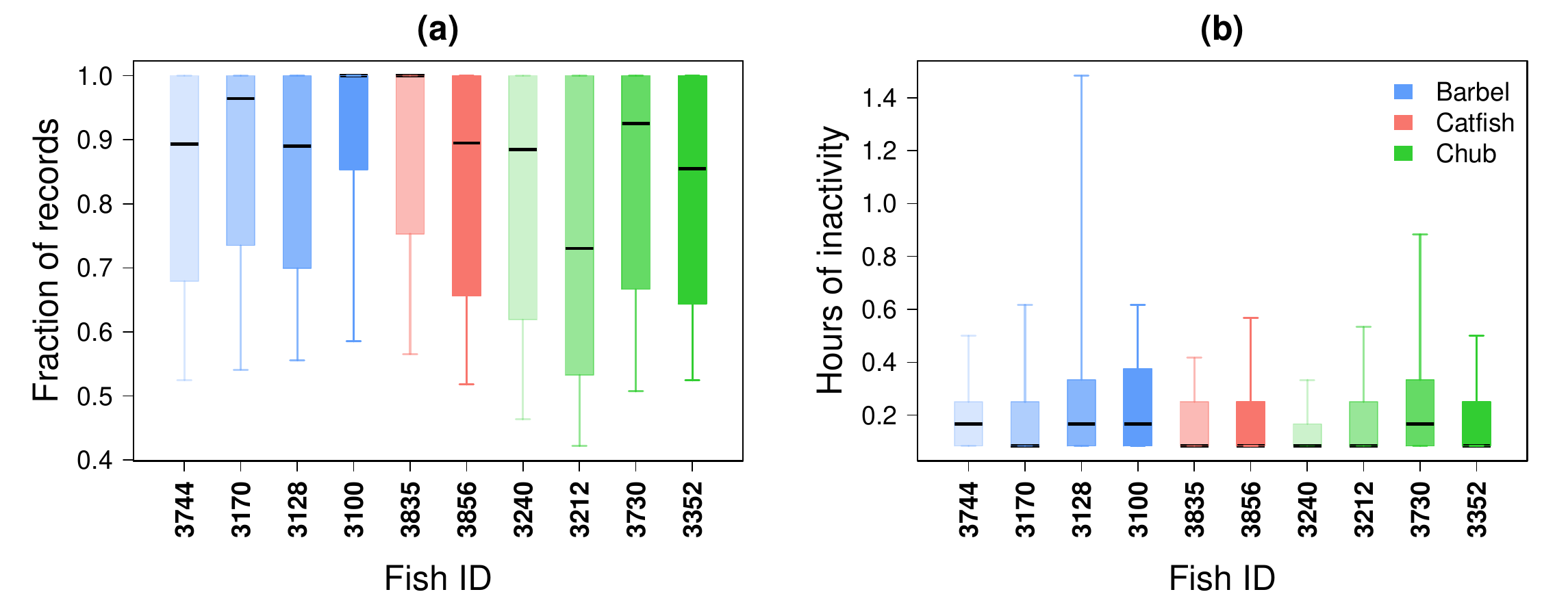}
	\caption{\textbf{Daily fish trajectory reconstruction accuracy.} (a) Boxplots of the fraction of records observed in the location with the highest number of records during a 5-minute period. (b) Boxplots of the duration (in hour) of the sequence during which the presence of a fish is not recorded. The results have been aggregated over the ten selected days for each fish individual. Each boxplot is composed of the first decile, the lower hinge, the median, the upper hinge and the last decile. \label{FigS3}}
\end{figure}

\begin{figure}[!ht]
	\centering 
	\includegraphics[width=14cm]{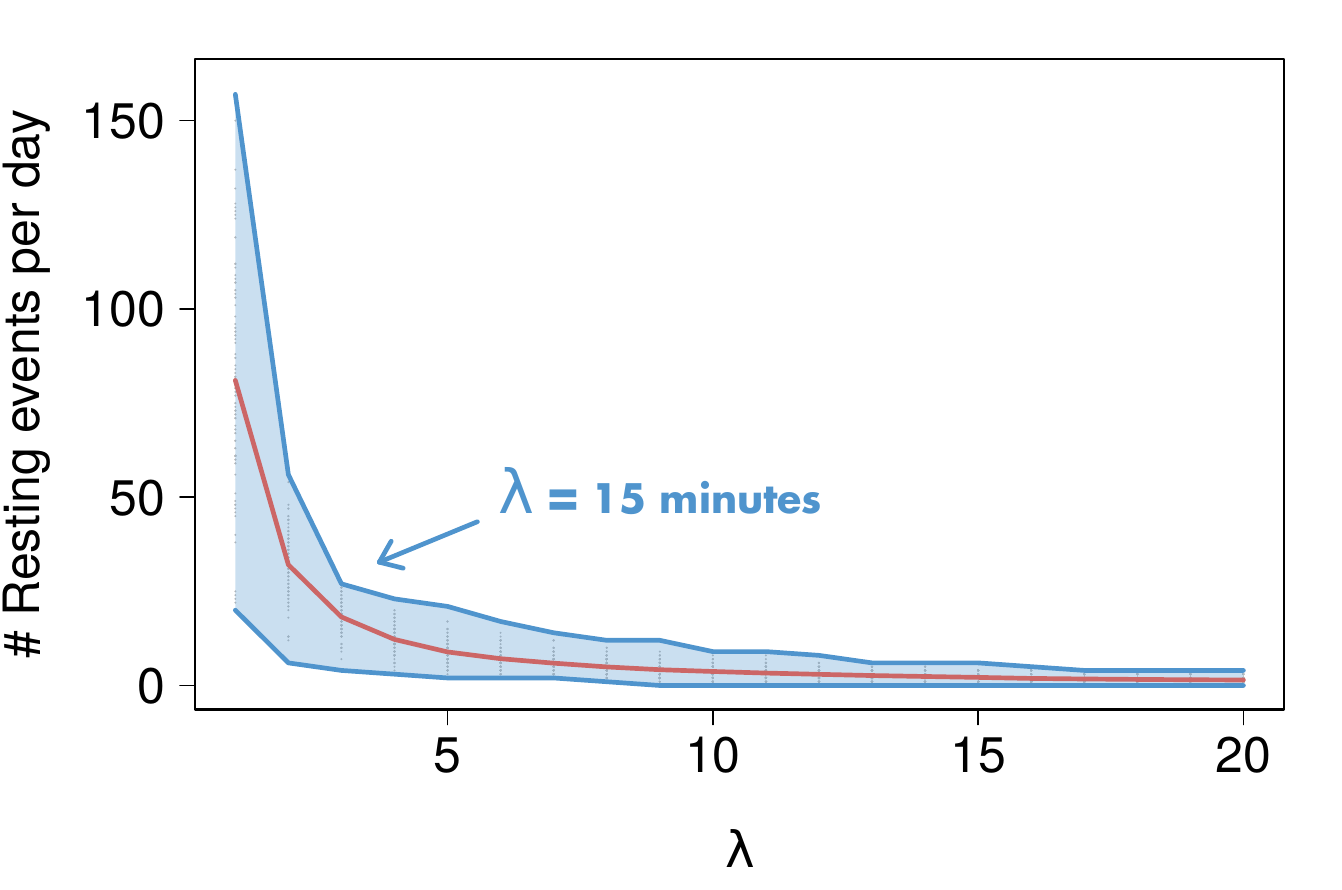}
	\caption{\textbf{Number of resting events per day as a function of $\lambda$.} The grey points represents the number resting events for the 100 daily spatio-temporal trajectories. The red line represents the average. The blue lines represent the minimum and maximum values. \label{FigS4}}
\end{figure}

\begin{figure*}[!ht]
	\centering 
	\includegraphics[width=12cm]{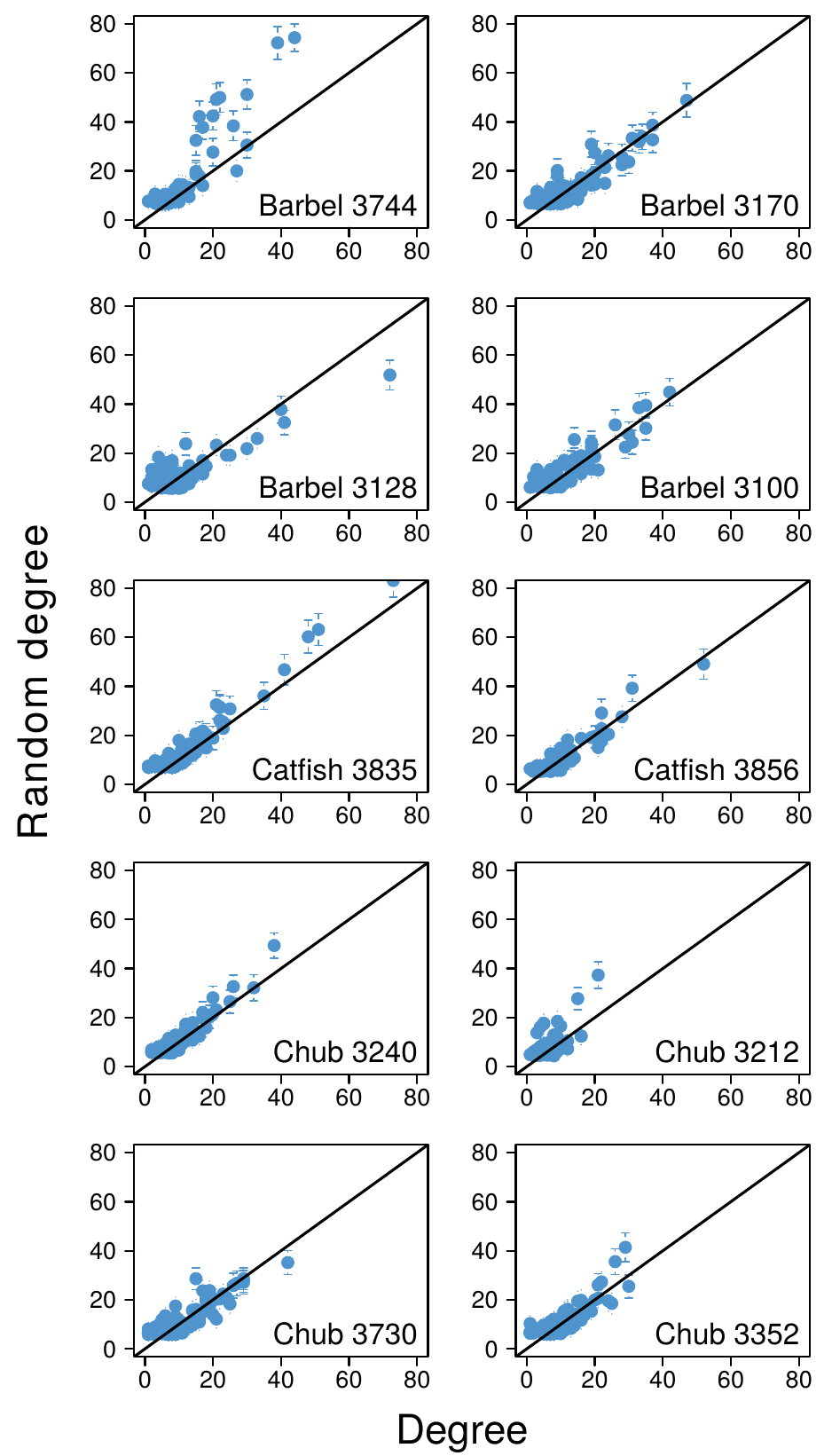}
	\caption{\textbf{Comparison between observed and average random degree distribution.} Each point represents an event with the observed average degree on the x-axis and the average degree obtained with the null model on the y-axis averaged over 100 replications (the error bars represent one standard deviation). \label{FigS5}}
\end{figure*}

\begin{figure*}[!ht]
	\centering 
	\includegraphics[width=16cm]{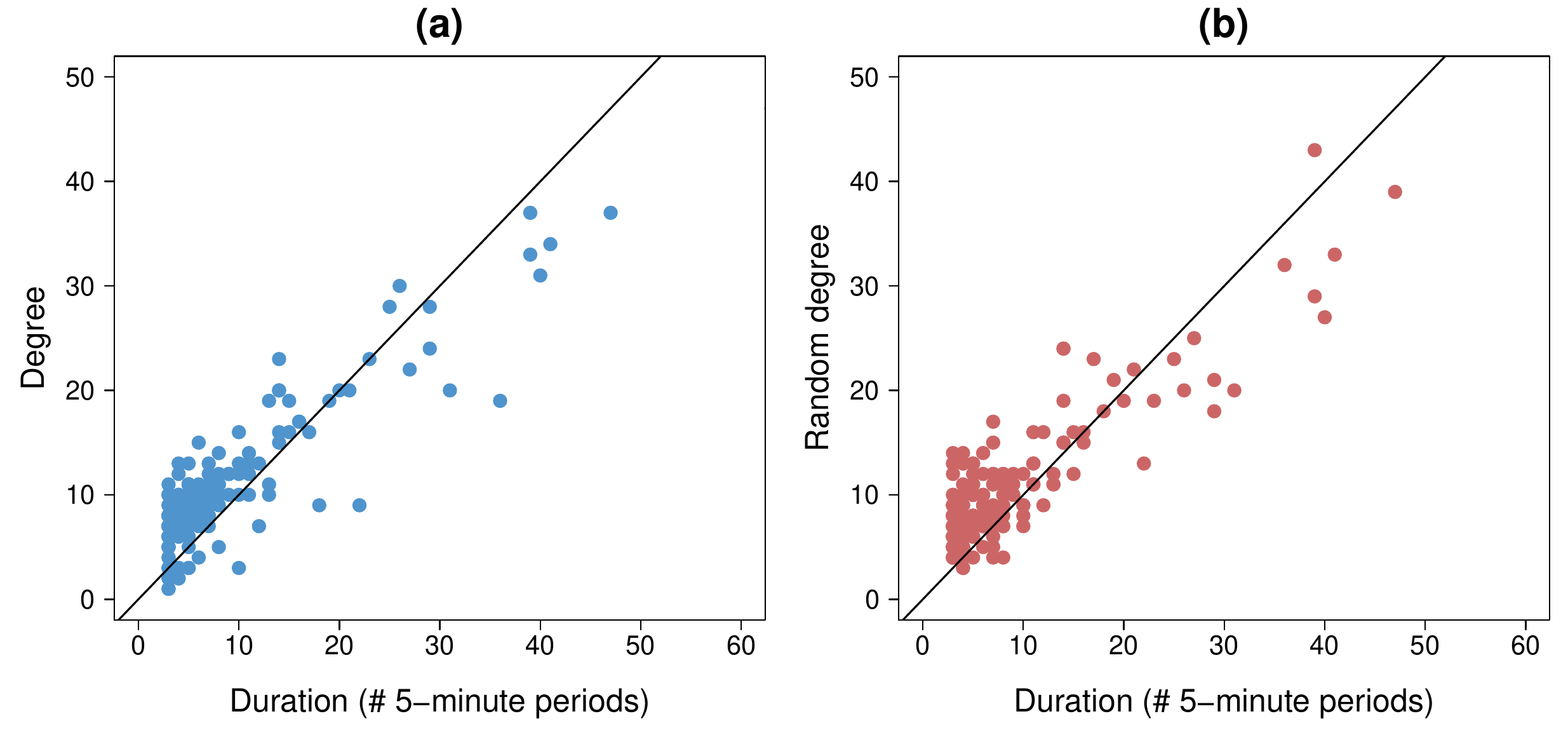}
	\caption{\textbf{Relationship between event degree and event duration.} Each point represents an event with its duration (in number of 5-minute periods) on the x-axis and the observed degree (a) and the one obtained with one realization of the null model (b)) on the y-axis averaged. \label{FigS6}}
\end{figure*}

\begin{figure*}
	\centering 
	\includegraphics[width=\linewidth]{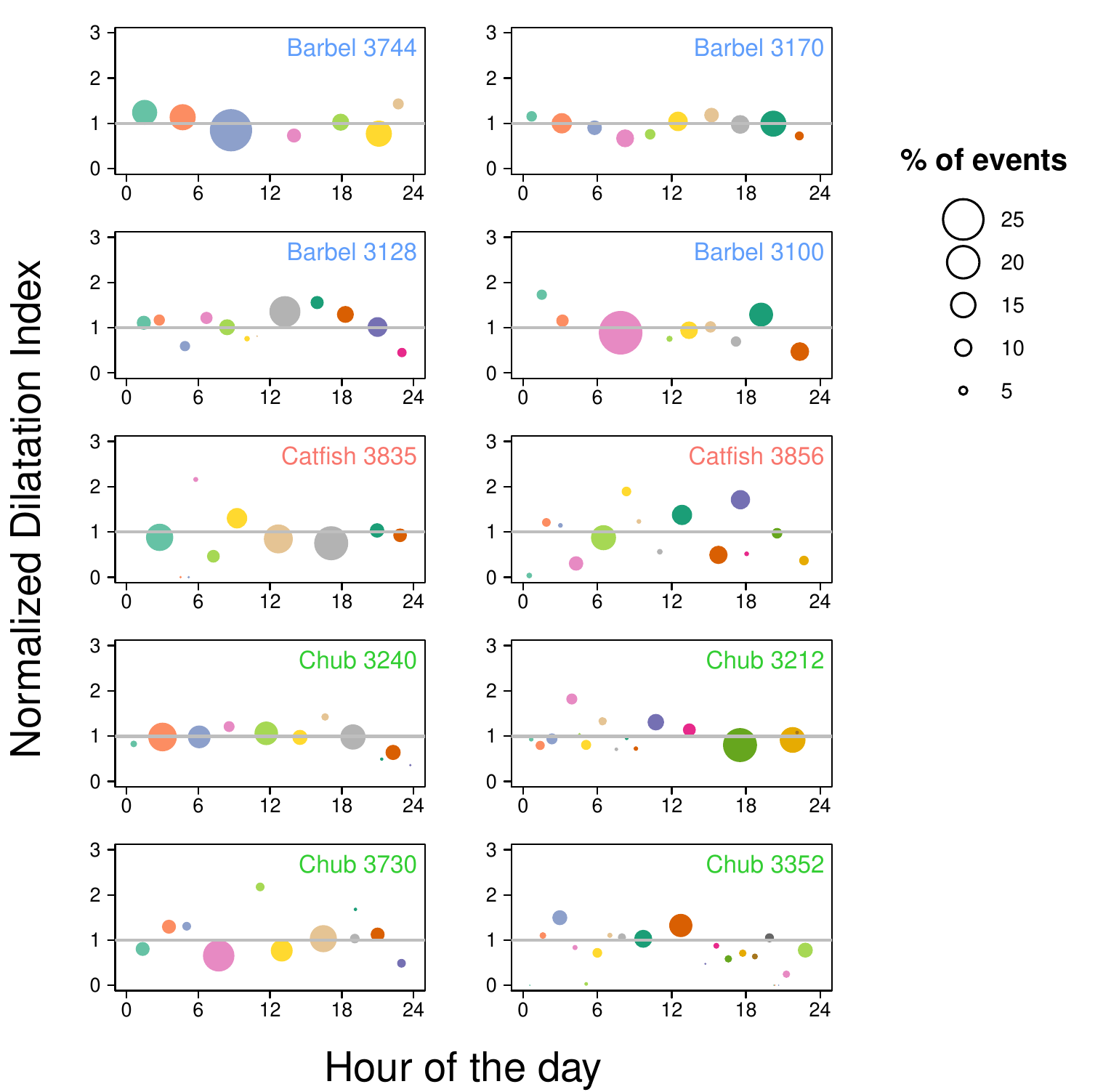}
	\caption{\textbf{Analysis of the event networks' community structure obtained with the null model.} The plots display the communities characteristics for every fish. Each point represents a community with the average resting time on the x-axis and the community dilatation index on the y-axis. The community dilatation index is normalized by the dilatation index $DI_{NM}$. The size of the dots is proportional to the fraction of events. Only results with one random network are shown. \label{FigS7}}
\end{figure*}

\newpage
\clearpage
\newpage
\section*{Supplementary Tables}

{\renewcommand{\arraystretch}{1.2}
	\begin{table*}[!ht]
		\caption{\textbf{Statistical properties of the resting event networks (standard deviations).} Number of events (\#Nodes), number of links (\#Links), average degree (Degree) and dilatation indices (expressed in meters). Standard deviations associated with the average values displayed in Table 1.}
		\label{TabS1}
		\begin{center}
			\begin{tabular}{cccccccccc}			
				\hline
				\textbf{Fish ID} & \textbf{Species} & \textbf{\#Nodes} & \textbf{\#Links} & \textbf{\#Links (NM)} & \textbf{Degree} & \textbf{Degree (NM)}  & \textbf{$DI$} & \textbf{$DI_{tot}$} & \textbf{$DI_{NM}$}\\
				\hline
				
				3744 & Barbel & NA & NA & 33.64 & NA & 0.42 & NA & NA & 20.14 \\ 
				3170 & Barbel & NA & NA & 33.25 & NA & 0.34 & NA & NA & 15.89 \\ 
				3128 & Barbel & NA & NA & 26.95 & NA & 0.31 & NA & NA & 18.67 \\ 
				3100 & Barbel & NA & NA & 31.39 & NA & 0.40 & NA & NA & 25.99 \\ 
				3835 & Catfish & NA & NA & 34.04 & NA & 0.36 & NA & NA & 11.97 \\ 
				3856 & Catfish & NA & NA & 23.64 & NA & 0.29 & NA & NA & 12.49 \\ 
				3240 & Chub & NA & NA & 26.56 & NA & 0.27 & NA & NA & 17.63 \\ 
				3212 & Chub & NA & NA & 21.39 & NA & 0.25 & NA & NA & 15.41 \\ 
				3730 & Chub & NA & NA & 31.86 & NA & 0.35 & NA & NA & 20.97 \\ 
				3352 & Chub & NA & NA & 32.30 & NA & 0.30 & NA & NA & 11.33 \\
				
				\hline			
			\end{tabular}
		\end{center}
	\end{table*}
}

{\renewcommand{\arraystretch}{1.2}
	\begin{table*}[!h]
		\caption{\textbf{Number of resting event network communities.} Standard deviations associated with the average values displayed in Table 2.}
		\label{TabS2}
		\begin{center}
			\begin{tabular}{cccc}			
				\hline 
				\textbf{Fish ID} & \textbf{Species} & \textbf{\#Com} & \textbf{\#Com (NM)}\\
				\hline
				
				3744 & Barbel & NA & 1.34\\ 
				3170 & Barbel & NA & 0.97\\ 
				3128 & Barbel & NA & 0.76\\ 
				3100 & Barbel & NA & 0.90\\ 
				3835 & Catfish & NA & 1.44\\ 
				3856 & Catfish & NA & 1.12\\ 
				3240 & Chub & NA & 0.85\\ 
				3212 & Chub & NA & 0.87\\ 
				3730 & Chub & NA & 1.11\\ 
				3352 & Chub & NA & 1.81\\ 
				
				\hline			
			\end{tabular}
		\end{center}
	\end{table*}
}

\end{document}